\documentclass[journal]{IEEEtran}

\usepackage{calc}

\usepackage{multirow}
\usepackage{cite}
\usepackage{graphicx,subfigure}
\usepackage{amsmath,amssymb}
\usepackage{color}
\usepackage{algorithm}
\usepackage{algorithmic}

\newtheorem{theorem}{Theorem}
\newtheorem{corollary}{Corollary}[theorem]

\newcommand{\bit}{\begin{itemize}}
\newcommand{\eit}{\end{itemize}}

\newcommand{\bc}{\begin{center}}
\newcommand{\ec}{\end{center}}

\newcommand{\ba}{\begin{array}}
\newcommand{\ea}{\end{array}}

\newcommand{\beq}{\begin{equation}}
\newcommand{\eeq}{\end{equation}}

\newcommand{\beqn}{\begin{equation*}}
\newcommand{\eeqn}{\end{equation*}}

\newcommand{\bean}{\begin{eqnarray*}}
\newcommand{\eean}{\end{eqnarray*}}
\newcommand{\bea}{\begin{eqnarray}}
\newcommand{\eea}{\end{eqnarray}}

\def\E{\mathbb{E}}

\newcommand{\Cc}{{\mathcal C}}

\newcommand{\Nc}{{\mathcal N}}

\begin{document}
\title{On-off Analog Beamforming for Massive MIMO}

\author{Shengli Zhang, \emph{Member, IEEE}, Chongtao Guo, \emph{Member, IEEE}, Taotao Wang, \emph{Member, IEEE}, and Wei Zhang, \emph{Fellow, IEEE}
	
\thanks{S. Zhang and C. Guo are with the Faculty of Information Engineering, Shenzhen University, Shenzhen, China (email: \{zsl, ctguo\}@szu.edu.cn). }
\thanks{T. Wang is with the Institute of Network Coding, The Chinese University of Hong Kong, Hong Kong, China, and the Faculty of Information Engineering, Shenzhen University, Shenzhen, China (email:  ttwang@ie.cuhk.edu.hk). }
\thanks{W. Zhang is with the School of Electrical Engineering and Telecommunications, the University of New South Wales, Sydney, Australia (email:  wzhang@ee.unsw.edu.au).
	}
}





\maketitle

\begin{abstract}
This paper investigates a new analog beamforming architecture for massive multiple-input multiple-output (MIMO) systems, where each of the multiple transmit antennas is switched to be on or off to form a beam according to the channel state information at transmitters. This on-off analog beamforming (OABF) scheme has the advantages in terms of both hardware complexities and algorithmic complexities. OABF can completely remove the high-cost, power-consuming and bulky analog phase-shifters that are extensively employed by traditional analog beamforming schemes; it only requires the deployment of low-cost analog switches that are easy to implement. Moreover, we show that the beams formed by such simple antenna on-off switch operations can achieve rather good performances with low-complexity beamforming algorithms. Specifically, we first propose two optimal signal-to-noise ratio maximization algorithms to determine the on-off state of each switch under the per-antenna power constraint and the total power constraint, respectively. After that, we theoretically prove that OABF can achieve the full diversity gain and the full array gain with complexities up to a polynomial order. Numerical results are consistent with our theoretical analysis. We believe that the simple structure of OABF makes massive MIMO much easier to implement in real systems. 
\end{abstract}


\section{Introduction}

Among the potential solutions for 5G cellular communications, massive multiple-input multiple-output (MIMO) has been shown to be able to increase the system spectral efficiency by several times \cite{rusek2013scaling, larsson2014massive}. Recently, extremely high frequency (EHF) bands from 30 to 300 GHz are designated for future wireless communications \cite{rangan2014millimeter}, since the available bandwidths over EHF are much wider than current cellular networks. For the deployment of wireless systems over frequencies higher than 30 GHz, a large amount of antennas are not only possible but also necessary to compensate the small array gain due to the compact antenna size \cite{analog_beamforming_2009, han2015large}.

A typical application of massive MIMO in cellular communications is that a base station (BS) equipped with a large number of antennas simultaneously serves mobile users using beamforming  \cite{gesbert2007shifting}. Digital beamforming is a well-developed technique for MIMO systems, because of its advantages of flexibility, adaptability, and performance optimality \cite{wiesel2008zero, gershman2010convex}. However, digital beamforming is too expensive and too power-consuming to be applied in massive MIMO systems, since it requires each antenna to be connected with one expensive radio frequency (RF) chain and a digital-to-analog converter (DAC) or analog-to-digital converter (ADC). The implementation of a large amount of RF chains and DACs/ADCs for digital beamforming is rather costly in massive MIMO systems.  

In order to save the number of RF chains and DACs/ADCs, analog beamforming is proposed and attracting more attentions. Analog beamforming usually employs RF phase shifters, variable gain amplifiers (VGAs) in the analog domain to form signal beams. The works on analog beamforming show that the performance gap of analog beamforming with respect to the conventional digital beamforming is justified by the reduction in its hardware costs \cite{zhang2005variable, sudarshan2006channel, analog_beamforming_ofdm_2010, analog_beamforming_channelest_2010, simplified_analog_beamforming_2011,liang2014low, hybrid_beamforming_2014, el2014spatially, sun2014mimo, alkhateeb2014mimo, alkhateeb2014channel, liu2014phase, han2015large, bogale2016number}.

Although analog beamforming has power and cost advantages over digital beamforming, the used RF phase shifters raise tremendous challenges and reliability issues for the design of RF hardwares, especially for millimeter wave (mmWave) carrier signals \cite{poon2012supporting}. First, RF phase shifters do not achieve fine enough phase resolutions, which limit the beamforming accuracy. Moreover, shifting signal phases in the RF domain degrades noise figures along with RF chains. Therefore, RF phase shifters make the implementation of analog beamforming still challenging and costly.

In this paper, we propose to further reduce the hardware complexity of analog beamforming by replacing all the bulky and costly RF phase shifters with RF switches. The RF switches are employed to control the on-off state of each transmit antenna according to the channel state information (CSI) at transmitters. We refer to this new analog beamforming architecture as on-off analog beamforming (OABF). The aim of OABF is to form beams over the air via these transmit antennas activated to be on state, without any phase and amplitude pre-processing neither in the analog domain nor in the digital domain. Actually, commercial RF switches have been widely used in wireless transceivers, and they possess very attractive properties such as cheap, compact size, fast speed, almost no power consumption, linear for wide bandwidth and high frequency \cite{switches_skyworks_2011}.\footnote{For example, according to the power model derived in \cite{switch_mimo_2016} from the recent progresses on mmWave RF circuits, typically the power of a RF phase shifter is round 30 mW and that of a RF switch is only around 5 mW.} Therefore, the OABF architecture is much easier for implementations than the RF phase shifters based analog beamforming. 

Intuitively, we can determine the on-off status of the switches of transmit antennas to maximize the received signal-to-noise ratio (SNR) --- we select the subset of antennas that have better channel conditions and similar phases for transmitting signals. At the first sight, the problem on finding the subset of transmit antennas that maximizes SNR looks like a combinatorial optimization problem, which is NP-hard in general. In contrast to this intuition, we find two optimal beamforming algorithms that determine the on-off status of each antenna for OABF with only a linear complexity and a polynomial complexity, respectively. Moreover, we theoretically prove that our OABF scheme can achieve the full  array gain and the diversity gain with the proposed beamforming algorithms. The main contributions of this paper are summarized as follows.   
\begin{enumerate}
	
	\item We propose OABF --- a new analog beamforming architecture that only relies on RF switches --- for massive MIMO systems. OABF can achieve beamsforming without any phase or amplitude  pre-processing neither in the analog domain nor in the digital domain. This simple OABF architecture further reduce the hardware complexity of analog beamforming. 
	
	\item We develop two optimal beamforming algorithms for OABF. The beamforming algorithms determines the on-off statues of the transmit antennas for maximizing the received SNR. One algorithm is developed under the per-antenna power constraint and it only has a linear complexity; the other is developed on the total power constraint and it only has a polynomial complexity. 
	
	\item We perform theoretical analysis to show that the simple OABF architecture can achieve the full array gain and the full diversity gain with the proposed beamforming algorithms; the achievable rate gap between the optimal phase-aligned beamforming and OABF is a constant $2{\log _2}\pi  = 3.3$ bits/symbol that dose not scale with the number of antennas and SNR. Simulations are performed to validate our results.
	
\end{enumerate}

\subsection{Related Works}

The analog beamforming algorithms and their corresponding analog RF architectures have been extensively studied in many works \cite{zhang2005variable, sudarshan2006channel, analog_beamforming_ofdm_2010, analog_beamforming_channelest_2010, simplified_analog_beamforming_2011,liang2014low, hybrid_beamforming_2014, el2014spatially, sun2014mimo, alkhateeb2014mimo, alkhateeb2014channel, liu2014phase, han2015large, bogale2016number}. For these analog beamforming schemes, the implementations of RF phase shifters and/or VGAs are necessary. 


In \cite{switch_mimo_2016}, the massive MIMO transceiver architecture that only employs RF switches is proposed. Its hardware structure is very similar to our OABF. However, there is no optimal solution to the beamforming problem found in \cite{switch_mimo_2016}. It claimed that the optimal beamforming problem has a complexity growing exponentially with the number of antennas, and suboptimal solution was given by using simultaneous orthogonal matching pursuit strategy \cite{el2014spatially}. 

Our OABF is related to the subset antenna selection schemes, where the best $K$ antennas are selected from the total $N$ antennas \cite{mmwave_combination_1996, antenna_selection_2004, antenna_selection_miso_2014}; however, OABF is substantially different from antenna selection. For the traditional antenna selection schemes, each selected antenna is connected with one RF chain (and followed by a phase shifting operation in the digital domain) to achieve a coherent combination. Therefore, $K$ is usually determined by the number of available RF chains, and beamforming effect comes from the phase shifting processing at the digital domain. For OABF, all the selected antennas are directly connected to one RF chain without any RF phase shifters in the analog domain and any phase shifting operations in the digital domain; the beamforming effect is simply achieved by selecting a subset of the antennas (see the later part for details).

The rest of this paper is organized as follows. Section II gives the architecture of OABF and the exploited system model. The beamforming algorithm under per-antenna power constraint and its performance analysis are presented in Section III. The beamforming algorithm under total power constraint and its performance analysis are presented in Section IV. In section V, numerical simulations are performed to verify the performance of OABF. Section VI concludes the paper.

\section{OABF Architecture and System Model}

\subsection{OABF Architecture}

In this part, we present the architecture of the proposed OABF. For a purpose of comparison, we first review three typical existing multi-antenna beamforming architectures, i. e., digital beamforming, phase-aligned analog beamforming and antenna selection \cite{linear_diversity_1959}.

\begin{figure}[!t]
	\centering
	\includegraphics[width=3.5in]{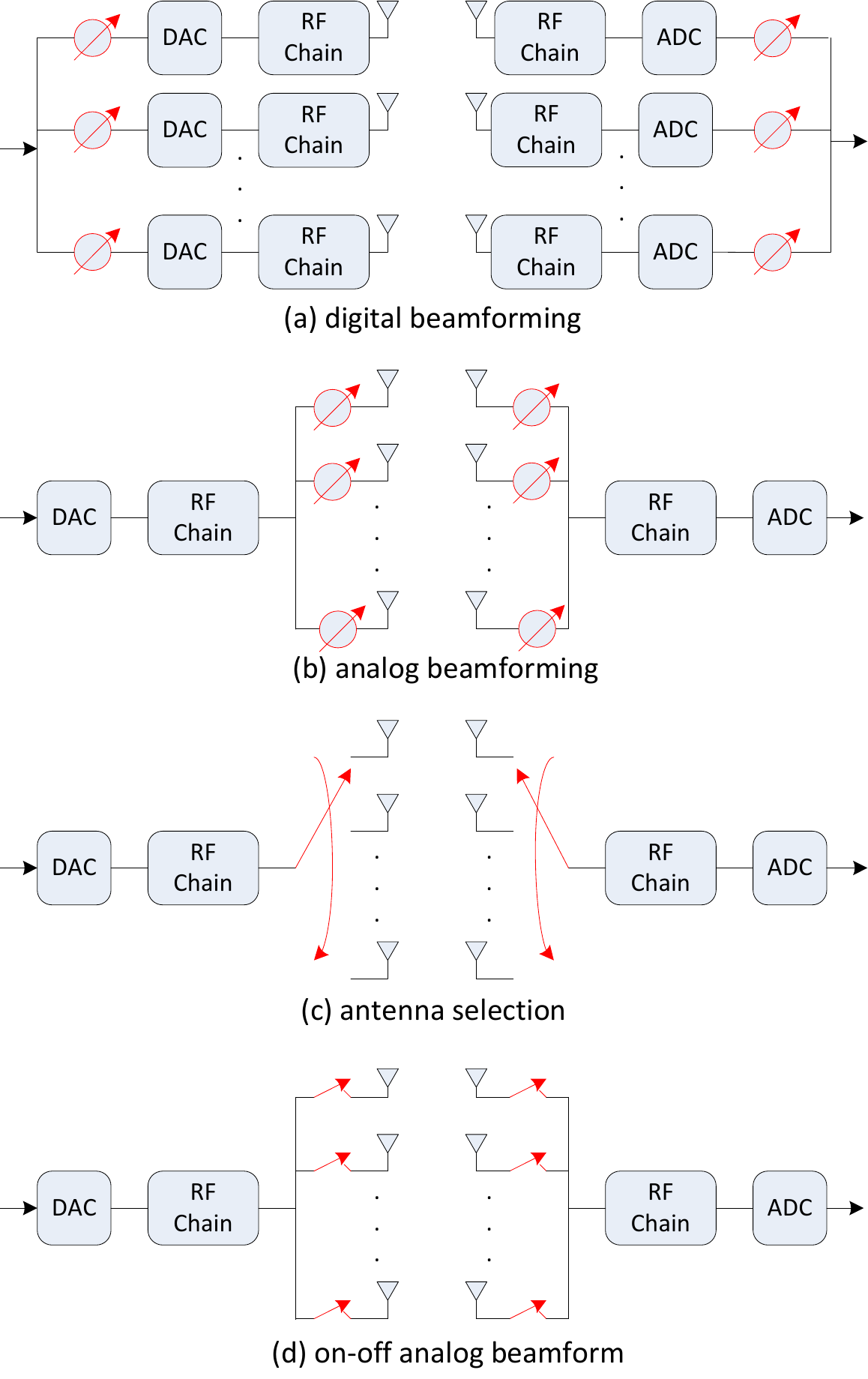}
	\caption{Block diagrams for the architectures of (a) digital beamforming; (b) analog beamforming; (c) antenna selection; (d) on-off analog beamforming. The left is the transmitter and the right is the receiver.} \label{beamforming_block_diagram}
\end{figure}

In most current wireless systems, fully optimal beamforming is implemented in the digital domain --- the signal phases (and amplitudes) are accordingly adjusted inside the digital signal processing (DSP) unit. The architecture of digital beamforming is illustrated in Fig. \ref{beamforming_block_diagram} (a)\footnote{For simplicity, this paper only considers the case of only one data stream transmitted from the BS to the mobile user. The extension to multiple data streams is straightforward. Without losing generality, we only consider beamforming by shifting signal phases, and we omit the amplitude adjusting operations.}, where the components highlighted by red are phase shifters. 

The advantage of the digital beamforming architecture is that the phase shifters can be easily implemented in the digital domain. This saves the expenses on RF circuits, especially for wireless systems over low frequency bands where phase shifters are difficult for design and very expensive for integration \cite{poon2012supporting}. The disadvantage of digital beamforming is that each antenna requires a RF chain and an ADC/DAC. This makes digital beamforming not suitable for massive MIMO systems. Moreover, for wireless systems with high frequency and wide bandwidth (e. g., mmWave systems), the costs of a RF chain, especially the price, power consumption and space cost, are much higher than that of a antenna \cite{poon2012supporting}; the implementation of the low power, high solution and fast DACs/ADCs are also very challenging \cite{poon2012supporting}. Therefore, we need alternative beamforming architectures for massive MIMO to save RF chains, DACs/ADCs and can also maintain the gains of multiple antennas.

One typical such scheme is the analog beamforming architecture that fulfills the beamforming operation in the analog domain, as illustrated in Fig. \ref{beamforming_block_diagram} (b). For analog beamforming, each antenna is connected with one RF phase shifter to rotate the signal phase in the analog domain. This analog beamforming is also called phase-aligned analog beamforming. The signal amplitudes can also be controlled in the analog domain using RF VGAs. In this paper, we just consider the RF phase shifter based phase-aligned analog beamforming as the benchmark to our OBFA. It is shown in \cite{linear_diversity_1959} that the same as digital beamforming, the phase-aligned analog beamforming can achieve the full diversity gain and the full array gain, which equal to the antenna number $N$. 

However, analog RF phase shifters, especially these working with wide bandwidth and high frequency, are also very costly and bulky \cite{poon2012supporting}. A further lower complexity/cost architecture is antenna selection, where only one antenna is selected to connect to one RF chain, as shown in Fig. \ref{beamforming_block_diagram} (c). Antenna selection schemes can also obtain the full diversity gain \cite{linear_diversity_1959}. But, its array gain is only $\sum_{i=1}^N 1/i$, which is much smaller than the full array gain of $N$. Antenna selection is then extended to select $K$ antennas from all the antennas if there are $K$ available RF chains. For a multiple-input single-output transmission, simply selecting the $K$ channels with largest channel gains is already the optimal  antenna selection scheme.

With a comparison between analog beamforming and antenna selection, we can find that analog beamforming utilizes all the antennas to align signal phases; while antenna selection only utilizes one antenna (connected with one RF chain) without any further analog signal processing. In this paper, we propose a new low complexity analog beamforming architecture in between antenna selection and analog beamforming. The proposed analog beamforming architecture is OABF (as shown in Fig. \ref{beamforming_block_diagram} (d)), where a subset of the antennas is selected to connected to one RF chain and these antennas are used to transmit signals without any analog signal processing; other antennas are disconnected. In our OABF, the RF switches on the selected antennas are kept on while others are kept off. Therefore, we refer to it as on-off analog beamforming (OABF).

The intuition for our OABF is that a subset of the $N$ antennas, with better channel conditions and similar phases, is selected to connect to the RF chain and thus can form a signal beam with large SNR. The connections between OABF and analog beamforming/antenna section are as follows. If the cardinality of the subset is limited to $1$, OABF degrades to the antenna selection scheme; if the analog coefficients in analog beamforming is limited to be $0$ or $1$ --- RF phase shifters are replaced by RF switches, it reduces to our OABF.

\subsection{Baseband System Model}

This part introduces the baseband system model and its relative notations for OABF. Without loss of generality, this paper considers a point-to-point transmission with $N$ antennas at the transmitter and one antenna at the receiver. The extension to multiple receive antennas is straightforward. We assume that there is only one data stream, hence one RF chain at each side. The channel between the $j$-th transmit antenna and the receive antenna is denoted by $h_{j}$, which are perfectly known at the transmitter by some feedback schemes or the reciprocity of the channel. We further assume that all $h_{j}, 1\leq j \leq N$ are independent identically distributed variables with complex Gaussian distribution, i.e., $h_j \sim \Cc\Nc(0,1)$ for all $j$. The channel coefficients keep constant during one packet transmission and change independently between different packet transmissions. This implies a block
fading environment that is valid for indoor channels and many outdoor channels \cite{rappaport1996wireless}.

\begin{figure} [t]
	\centering
	\includegraphics[width=3.5in]{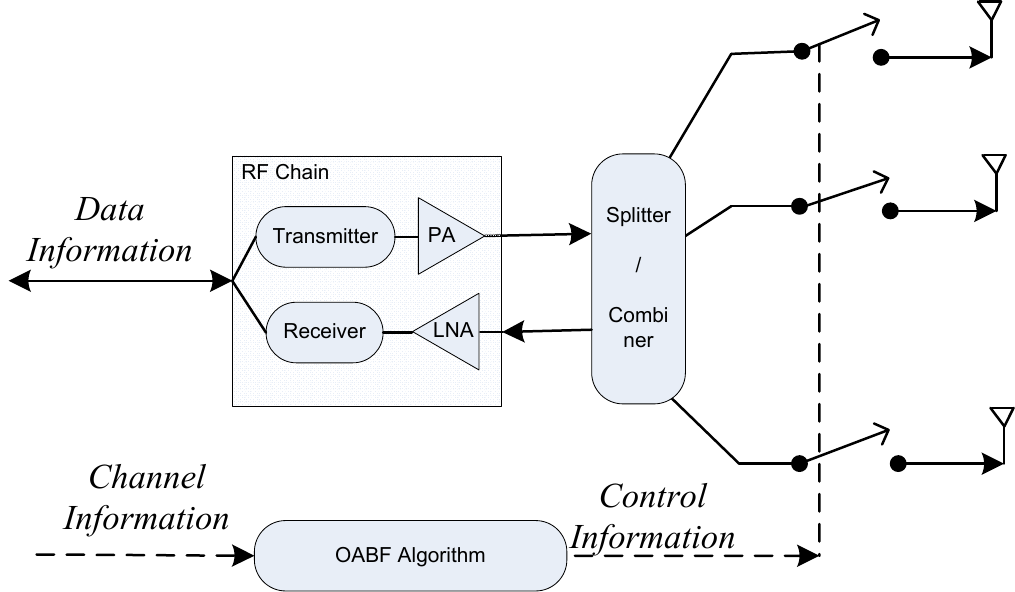}
	\caption{The system structure for on-off analog beamforming at the transmitter.}
	\label{fig_system_model}
\end{figure}

A more detailed OABF transmit architecture is illustrated in Fig. \ref{fig_system_model}. We denote the transmit power for the $j$-th antenna by $P_j$. At the transmitter side, some antennas that forms a set $T$ are selected for transmissions. Then, the baseband received signal can be expressed as
\begin{align} \label{eq:system_model}
y &=  \sum_{h_j \in T} \sqrt{P_j}h_{j} x  + n \nonumber \\
&= \sum_{i=1}^N \sqrt{P_i}I_i h_{i} x  + n  \nonumber \\ 
& \ \ \ \  s.t. \ I_i=1 \ if \ h_i \in T, \ \ else \ I_i=0 
\end{align}
where $n \sim \Cc\Nc(0,\sigma^2)$ is the receiver side additive white Gaussian noise and $I_i$ is an indicate variable, $T$ is a subset of the set $\{ h_1,h_2 \cdots h_N \}$. By intuition, select the best subset $T$ to optimize the received SNR is a combination optimization problem and has an exponential complexity with respect to the number of antennas, as discussed in \cite{switch_mimo_2016}. 

In this paper, we consider two cases of power constraint, the per-antenna power constraint,  i.e., $P_j \leq P_o, \forall h_j\in T$, and the total transmit power constraint i.e., $\sum_{h_j \in T} P_j \leq P_t$. For both cases, we develop beamforming algorithms with low complexities.




For performance comparisons, we analyze two basic asymptotic gains of beamforming: the array gain and the diversity gain. As defined in \cite{array_gain_2000}, the array gain is referred to as the increase of the average SNR achieved by using multiple antennas, with respect to the single-input single-output case; the diversity gain is referred to as the decrease ratio of the bit error rate $P_e$ averaged over the fading,  $d=-\lim_{SNR \to \infty} \frac{\log P_e}{\log(SNR)}$.

\section{OABF with Separate Power Constraint}
In this section, we consider the scenario that the power of each transmit antenna is separately constrained by $P_o$ and there is no constraint on the total transmit power.  
This assumption results from the case that each antenna in a practical transmitter is driven by a separate power amplifier that operates properly only when its transmit power is below a predesigned threshold. For current analog beamforming schemes, the separate transmit power constraint is sometimes more relevant than the total power constraint \cite{yu_per_antenna_power}.

Then, each selected antenna will transmit with the maximized power $P_o$ and the system model in \eqref{eq:system_model} can be rewritten as
\beq \label{eq:system_model1}
y =  \sqrt{P_o}\sum_{j \in T}h_{j} x  + n.
\eeq
Therefore, the received SNR at the receiver can be expressed as
\beq \label{SNR_case1}
SNR_{s} = \frac{P_o | \sum_{j \in T} h_{j} |^2} { \sigma^2}.
\eeq

In this section, we first introduce an optimal and linear complexity algorithm to determine the set $T$ that maximizes $SNR_s$ for the separate power constraint case. That beamforming algorithm is referred to as OABF-s. After that, we prove the full diversity gain and the full array gain of OABF-s.  



\subsection{Optimal OABF-s Algorithms}
With the separate power constraint on each antenna, maximizing $SNR_{s}$ is equivalent to maximizing the received signal power $| \sum_{j \in T} h_{j} |^2$  in \eqref{SNR_case1}. Specifically, we can find the optimal subset $T^*$ by 
\beq \label{opt_problem}
T^*= \arg \max_{ T } |\sum_{j \in T} h_{j}|
\eeq
At a first glance, the problem looks like a combinatorial optimization problem, which is NP hard in general and the high complexity is not affordable for massive MIMO. However, we will show that this optimization problem in \eqref{opt_problem} is actually not NP hard by proposing an optimal algorithm only with a linear complexity. An intuitive explanation is that channel coefficients $h_j$ in $T^*$ must have close phases to contribute to each other, and we only need to consider the sets whose elements have adjacent phases. The detailed algorithm is elaborated as follows.

\begin{figure}[t]
	\centering
	\includegraphics[width=3.5in]{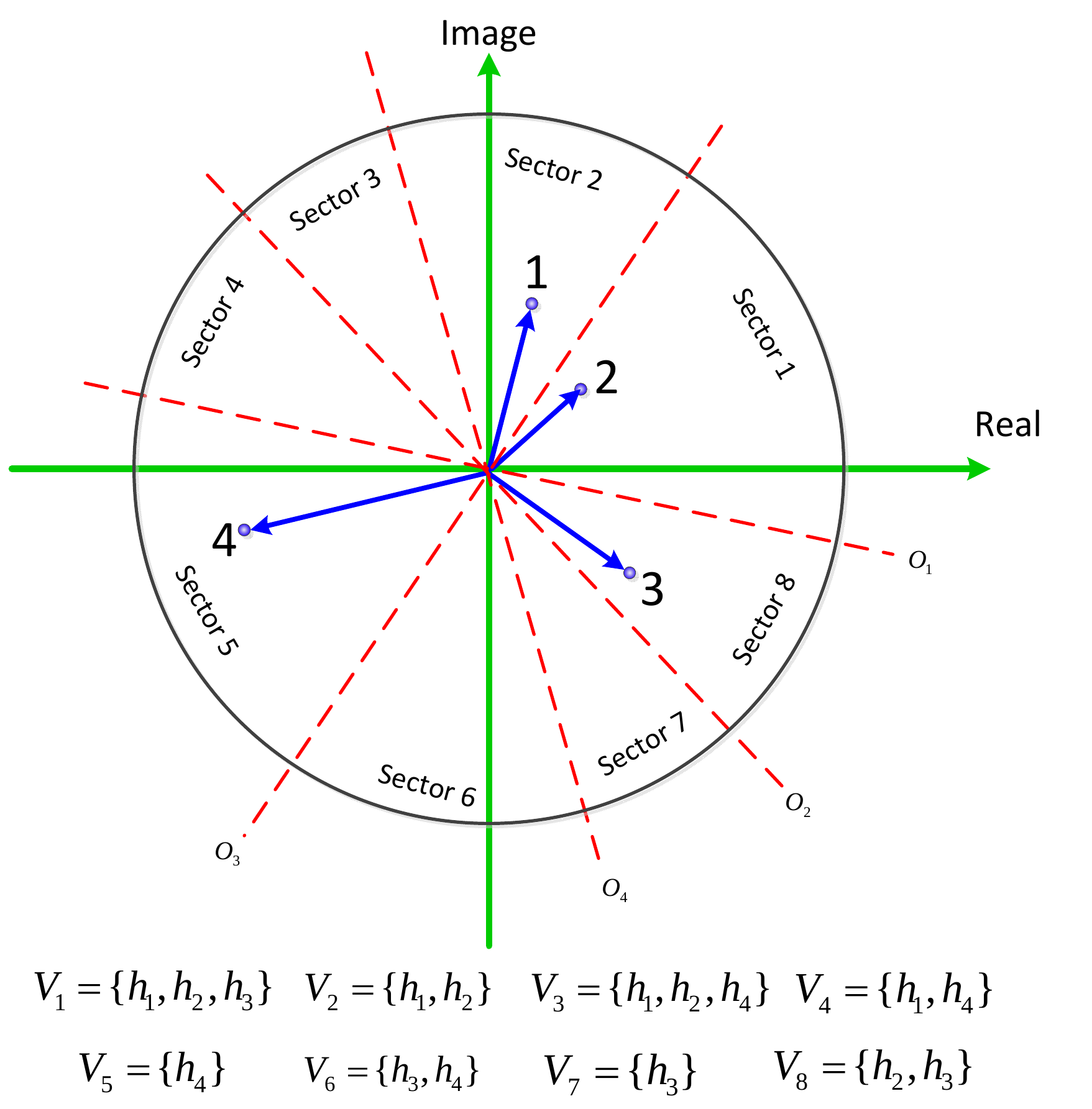}
	\caption{The complex plane where the four complex channel coefficients are denoted by the blue vectors; their orthogonal lines are denoted by red dash lines and they divide the whole plane into eight sector regions.}
	\label{fig_algorithm1}
\end{figure}

We first plot all the $N$ complex channel coefficients $h_j$ ($j=1,2,\cdots, N$) on a two-dimensional complex plane, where the horizontal and vertical axes correspond to their real and imaginary components, respectively. Since every complex number can be mapped to a vector on the complex plane, we use a complex number and its vector (on the complex plane) interchangeably in our descriptions, when ambiguity is not caused.

For each vector that represents a complex channel coefficient $h_j$, we plot its orthogonal line $O_j$ passing through the origin on the complex plane, $j=1,2,\cdots, N$. Then, the complex plane is divided into $2N$ sectors by the $N$ orthogonal lines. An example is shown in Fig. \ref{fig_algorithm1}, where there are $4$ complex channel coefficients (their vectors are denoted by blue solid lines) and their orthogonal lines (denoted by red dot lines) divide the complex plane into $8$ sectors.

As shown in the example of Fig. \ref{fig_algorithm1}, each orthogonal line $O_j$ bisects the complex plane into two halves; for any vector lying in the same half plane as a channel coefficient $h_j$, the channel coefficient $h_j$ has positive projection on this vector, and vice versa. Since each sector must be in one of the two half planes divided by each $O_j$, we can observe an important property of the sectors: the projections of a channel coefficient $h_j$ on any vectors that fall in the same sector always have the same sign, i.e., minus or plus.
For sector $k$ ($k=1,2,\cdots, 2N$), we use $V_k$ to denote the set of complex channel coefficients where each element $h_j$ lies in the same half plane divided by $O_j$ as sector $k$ --- all the channel coefficients in set $V_k$ have positive projections on the vectors falling in sector $k$. In Fig. \ref{fig_algorithm1}, we also show these sets with respect to each sector.

We define that $f^*$ is the summation of all the elements in the optimal set $T^*$, i.e., $f^*=\sum_{h_j\in T^*}h_j$. Then, we have the following result for $f^*$, which will help us to determine the optimal set $T$.

\begin{theorem}
If $f^*$ is located in sector $k$, then we have the optimal set $T^*=V_k$.
\end{theorem}

\begin{IEEEproof} It can be proved by contradictions to the assumption $T^* \neq V_k$. On one hand, if there is any complex number $h_i$ in $V_k$ but not in $T^*$, we can increase the value of $|\sum_{h_j\in T^*}h_j|$ by adding $h_i$ to $T^*$, since $h_i$ has a positive projection on $f^*$ ($f^*$ falls in sector $k$). On the other hand, if there is any complex number $h_i$ not in $V_k$ but in $T^*$, we can increase the value of $|\sum_{h_j\in T^*}h_j|$ by deleting $h_i$ from $T^*$, since $h_i$ has a negative projection on $f^*$. Therefore, both cases find contradictions to the assumption that $V_k$ is not the optimal set. Hence our theorem is proved.
\end{IEEEproof}

Due to the fact that $f^*$ must fall in one of the $2N$ sectors, the optimal set $T^*$ must equal one of the $2N$ sets $V_k$ for $k=1,2,\cdots,2N$. In other words, the optimal set $T^*$ can be determined as $T^*=V_{k^*}$, where $k^*=\arg\max_{k}|\sum_{h_j\in V_k}h_j|$. And a consequence of {\it {Theorem 1}} is the statement in the next
corollary.
\begin{corollary}
When changing from one set $V_{k-1}$ to the next set $V_k$, at least one elements in $V_{k-1}$ is removed or at least one new element is added. (The proof of this corollary is straightforward and we omit it here.
)
\end{corollary}

According to the above discussion, we can summarize the algorithm for obtaining the optimal $T^*$ as the following algorithm.

\begin{algorithm}[H]
\caption{OABF-s  Algorithm} \label{oabfs_algorithm}
\begin{algorithmic} [1]

\STATE Partition the whole space in to $2N$ sectors based on the orthogonal lines of the $N$ complex channel coefficients.

\STATE For each sector $k$, we determine the associated set $V_k$. Particularly, if $h_i$ has a positive projection on sector $k$, then we have $h_i \in V_k$; otherwise, we have $h_i \notin V_k$.
\STATE Calculate $f_1=\sum_{j\in V_1}h_j$.

\STATE Calculate $f_k= f_{k-1}-\sum_{j\in V_{k-1}\setminus V_k} h_j+\sum_{j \in V_{k}\setminus V_{k-1}}h_j$ for all $k=2, 3, \cdots, 2N$.
\STATE Obtain $k^* = \arg \underset{k=1,2,\cdots, 2N}{\max} |f_k|$ and obtain the corresponding set $T^*$ by $T^*=V_{k^*}$.

\end{algorithmic}
\end{algorithm}

%
%

\subsection{Performance Analysis}
In this part, we show that the OABF-s algorithm can lead to the global optimality with linear-time complexities. We also prove that the OABF architecture with the OABF-s algorithm can achieve the full diversity gain and the full array gain.

\subsubsection{Complexity} As shown in Algorithm \ref{oabfs_algorithm}, each channel coefficient is added for one time and minus for another time (only a  part of overlapped channel coefficients in $V_1$ are involved in the calculation four times), and the complexity is linear with the number of antenna elements. Specifically, the algorithm can lead to the global optimality with a linear complexity $\mathcal{O}(N)$. 

\subsubsection{Diversity Gain} Our OABF-s scheme performs better than the antenna selection scheme, since antenna selection is only a special case of our OABF-s. Since antenna selection can achieve the full diversity gain, OABF-s must also be able to achieve the full diversity order $N$.

\subsubsection{Array Gain} For the array gain analysis, we first present a sub-optimal algorithm, referred to as OABF-b, to bridge the optimal OABF-s and the antenna selection scheme. We will then theoretically prove that  OABF-b can achieve the full array gain. The sub-optimal OABF-b algorithm is given as follows.

\begin{algorithm}[H]
\caption{OABF-b  Algorithm} \label{map_algorithm1}
\begin{algorithmic} [1]
   \STATE  Select the channel coefficient with maximum amplitude denoted as $h_m, m=\arg\max_{i=1,2,\cdots, N}  |h_i|$
   \STATE  Make a set $R=\{h_m\}$.
   \STATE  Add all elements $h_i$ to $R$ if its projection on $h_m$ is positive, i.e., $R=\{ R, h_i \}$ if $ (h_i*h_m+h_i^{*}*h_m^{*}) \ge 0$.
\end{algorithmic}
\end{algorithm}

With reference to Fig. \ref{fig_algorithm1}, we can see that the OABF-b algorithm simply divides the complex plane into two halves by the orthogonal line of $h_m$; then it selects all the complex channel coefficients at the same half plane as $h_m$. We have the following theorem for OABF-b.
\begin{theorem}
The OABF-b algorithm can achieve the full diversity gain and the full array gain under the per-antenna power constraint.
\end{theorem}
\begin{IEEEproof}
We prove the full array gain by proving that the expected received SNR of OABF-b is larger than that of the phase-aligned analog beamforming scheme divided by a constant factor $\pi^2$. The detailed proof can be found in the appendix.
\end{IEEEproof}
	
Since the optimal OABF-s algorithm performs better than OABF-b, OABF-s can also achieve the full array gain with a linear complexity.

Since the SNR loss of OABF-b is less than $1/\pi^2$ compared to the optimal phase-aligned analog beamforming, we can directly obtain the upper bound of the rate loss for the OABF-s algorithm. We state it in the following corollary.
\begin{corollary}
Compared to phase-aligned analog beamforming, the achievable rate loss of OABF-s is upper bounded by $2\log_2 \pi$, which is independent of the antenna number.
\end{corollary}

\subsubsection{Robustness}
Although we assume perfect CSI at the transmitter, our OABF-s algorithm is robust to inaccurate CSI. First, OABF-s only depends on the relative phase of the channel coefficients, the channel amplitude errors does not affect the function of OABF-s. Moreover, since we have proved that OABF-b can also achieve the full diversity gain and the full array gain, a suboptimal algorithm (can be treated as a version of the optimal algorithm with errors) can also achieve a near optimal performance.

\section{OABF with Total Power Constraint}
So far, we have studied the OABF algorithm subject to the per-antenna power constraint at the transmitter.
In this section, we study the case where a total power constraint $P_t$ is evenly allocated to all the $K$ selected antennas. This power constraint  means that we just need to deploy one power amplifier at the RF chain.

With the total power constraint $P_t$, the transmit power of each antenna is $P_t/K$ and the system model in \eqref{eq:system_model} can be rewritten as
\beq \label{eq:system_model2}
y =  \sqrt{P_t/K}\sum_{h_j \in T}h_{j} x  + n.
\eeq
where $K$ is the cardinality of the set $T$.
The received SNR can then be expressed as
\beq \label{SNR_case2}
SNR_{t} = \frac{P_t | \sum_{j \in T} h_{j} |^2} { K\sigma^2}.
\eeq
From \eqref{SNR_case2}, we can see that maximizing $SNR_t$ is equivalent to maximizing $\frac{| \sum_{j \in T} h_{j} |^2} { K}$. Different from the case of per-antenna power constraint, we need to find a set not only with a large summation but also with a small cardinality.  

In this section, we first introduce the optimal algorithm, referred to as OABF-t, to determine the optimal set $T^*$ that maximizes $\frac{| \sum_{j \in T} h_{j} |^2} {K}$. After that, we prove the full diversity gain and the full array gain of OABF-t by showing its superiority over OABF-s.

\subsection{Optimal OABF-t Algorithm}
For the previous OABF-s algorithm, the essential idea is to combine a number of complex channel coefficients $h_j$ to obtain one set $V_k$ such that all the elements of this set have positive contributions (projections) to any vectors in sector $k$. And all the elements outside this set have non-positive contributions to any vectors in sector $k$. In other words, given one sector $k$, there is an unique such set $V_k$.
Extending this idea, we obtain the effective part of OABF-t algorithm, which maximizes  $| \sum_{h_j \in T} h_{j} | $ under the condition that the cardinality of the set $T$ is given. Then, the OABF-t algorithm, which  maximizes $\frac{| \sum_{j \in T} h_{j} |^2} { K}$, is finally obtained by checking all possible cardinalities of the sets.

First, we use the obtained CSI to find the set $T$ that maximizes $| \sum_{h_j \in T} h_{j} | $ and the cardinality of set $T$ (the number of selected antennas) is $K$. We use $T^{(K)}$ to denote this set with $K$ elements that maximizes $|\sum_{h_j \in T^{(K)}} h_{j} |$. We use $\theta^{(K)}$ to denote the phase of $f^{(K)}=\sum_{h_j \in T^{(K)}} h_{j}$. Then, we sort the $N$ channel coefficients $h_j$ in the descending order according to their projections on $f^{(K)}$, denoted as $h_{\pi_j}$. Then $T^{(K)}$ must be formed by the first $K$ coefficients, i.e., $h_{\pi_1}, h_{\pi_2}, \cdots, h_{\pi_K}$. We refer to this process as the generation of set $T(\theta^K,K)=\{h_{\pi_1}, h_{\pi_2}, \cdots, h_{\pi_{K}}\}$ based on the vector angle $\theta^{(K)}$.

Second, we see that it is possible to divide the whole complex plane, angle from $0$ to $2\pi$, into $M$ sectors, which satisfies the following conditions: i) for any $\theta$ in the same sector, all generated sets $T(\theta, K)$ are always the same; ii) for any $\theta'$ outside this sector, we will generate a different set $T(\theta', K)$; iii) for any two adjacent sectors, their corresponding sets only differ by one element. These conditions can be explained by noting that the values of the $N$ channel coefficients are discrete. Therefore, there must be a small sector with angles $[\theta-\delta_l, \theta+\delta_r]$, determined by two small values $\delta_l, \delta_r$, such that $T(\theta, K)$ is the same for on any $\theta$ in this sector, $T(\theta, K)$ changes when $\theta$ is outside this range. Since the change of $\theta$, as well as $h_j$'s projection on it, is continuous,  there must be an adjacent sector,  $[\theta+\delta_r, \theta+\delta_{r_1}]$, such that for any $\theta'$ in it we will generate a new set $T(\theta', K)$ with only one different element from $T(\theta, K)$.

Third, we divide the whole complex plane into such proper sectors by finding the boundaries between adjacent sectors. At a boundary, the two different elements of the two corresponding sets should have the same projections. In particular, for a given $\theta$ which belongs to one sector, $T(\theta, K)=\{h_{\pi_1}, h_{\pi_2}, \cdots, h_{\pi_{K}}\}$ is the set formed with the largest $K$ projections, and $T^C=\{h_{\pi_{K+1}}, h_{\pi_{K+2}}, \cdots, h_{\pi_{N}}\}$ is the complementary set formed by the other channel coefficients. The different element, belonging to the set of the next sector, must also belong to $T^C$. To find the boundary, we need to find a minimum phase shift $\delta > 0$, such that
\beq
\min_{h_i \in T} |h_i|\cos{(\theta+\delta-\theta_i)} = \max_{h_j \in T^C} |h_j|\cos{(\theta+\delta-\theta_j)},
\eeq
where $\theta_i$ and $\theta_j$ are the phase of $h_i$  and $h_j$, respectively. Therefore, $\delta$ can be calculated as
\beq \label{eq:delta_calculation}
\delta = \arctan \left( \min_{h_i \in T,h_j \in T^C} \frac{|h_j|\cos(\theta-\theta_j)-|h_i|\cos(\theta-\theta_i)}{|h_i|\sin(\theta-\theta_i)-|h_j|\sin(\theta-\theta_j)} \right).
\eeq
Then, $\theta+\delta$ is a boundary of the sector. In a similar way, we can find all the boundaries so as to determine all the sectors.

As discussed above, we can generate the set $T^K$ for each sector and find the maximum value $|\sum_{h_j \in T^{(K)}} h_{j} |$ of all the generated sets. The detailed algorithm of OABF-t is given as Algorithm \ref{oabt_algorithm}.

\begin{algorithm}[!h]
	\caption{OABF-t  Algorithm} \label{oabt_algorithm}
	\begin{algorithmic} [1]
		
		\STATE Calculate the set $T^s$ with OABF-s Algorithm and $K^s=|T^s|$.
		
		\STATE Initialize $K^*=1$,$f^*=\max_{i} |h_i|^2$ and $T^*={h_j}$ where $j=\arg\max_{i} |h_i|$.
		
		\FOR{$K=2$ to $K^s$}
		\STATE Initialize $\theta=0$, $f=0$.
		\STATE Sort $h_j$ according to their projections on $\theta$, denoted as $h_{\pi_j}$
		\STATE Generate two sets as $T=\{h_{\pi_1}, h_{\pi_2}, \cdots, h_{\pi_{K}}\}$ and $T^C=\{h_{\pi_{K+1}}, h_{\pi_{K+2}}, \cdots, h_{\pi_{N}}\}$
		
		\WHILE {$\theta < 2\pi$}
		\IF {$|\sum_{i=1}^{K} h_{\pi_i}|¡­¡­2 > f$}
		\STATE $f=|\sum_{i=1}^{K} h_{\pi_i}|^2$, $T^0=T$
		\ENDIF
		\STATE Calculate $\delta$ as in \eqref{eq:delta_calculation}
		\STATE Update $\theta$ as $\theta+\delta$,
		\STATE Update $T$ and $T^C$ by exchanging the two corresponding elements in \eqref{eq:delta_calculation}
		\ENDWHILE
		\IF {$f/K > f^*$}
		\STATE Update $ f^*=f/K$, $K^*=K$, $T^*=T^0$
		\ENDIF
		\ENDFOR
		
	\end{algorithmic}
\end{algorithm}




\subsection{Performance Analysis}
\subsubsection{Complexity} We first analyze the complexity of the OABF-t algorithm. In this algorithm, the number of the \emph{for} iterations is less than $N$; the number of the \emph{while} iterations is of the order $N$ since one element of the set $T$ is changed during one iteration; inside the \emph{while} iteration, the most complex processing is the calculation of $\delta$ in \eqref{eq:delta_calculation}, whose complexity is less than $N^2/4$. Therefore, the complexity of the OABF-t algorithm is of the order $N^4$, i.e., a polynomial complexity.

\subsubsection{Diversity gain and array gain} We now analyze the diversity gain and array gain of the OABF-t algorithm. Firstly, OABF-t performs better than the antenna selection scheme. Therefore it can achieve the full diversity gain. Secondly, OABF-t performs better than OABF-s. Therefore, based on the expected received SNR calculation in \eqref{eq:gamma_1_result}, the expected received SNR of OABF-t denoted by $\gamma_2$ can be bounded by that of OABF-s
\begin{align}
\gamma_2 &\ge \frac{P_t}{K^s \sigma^2} \left( N+N(N-1)\pi /4 \right) \\ \nonumber
& > \frac{P_t}{ \sigma^2} \left( 1+(N-1)\pi /4 \right).
\end{align}
It means that the order of the array gain that is the ratio of $\gamma_2$ over $P_t/\sigma^2$ for large $N$ can also achieve the largest value $N$. Thus, we have the following theorem.

\begin{theorem}
The optimal OABF-t algorithm can also achieve the full diversity gain and the full array gain with the total power constraint and a polynomial complexity.
\end{theorem}

\section{Numerical Results}
In this section, we present numerical results to show the performances of our OABF. In simulations, the noise variance at the receiver side is normalized to unit, i.e., $\sigma ^ {­2}=1$. For performance comparisons, we evaluate three different beamforming schemes: i) our OABF scheme; ii) the RF phase shifters based phase aligned analog beamforming scheme that is also referred to as the optimal scheme in this section; iii) the antenna section scheme. 

\subsection{OABF with Separated Power Constraint}

We first present the numerical results under the per-antenna power constraint. With a per-antenna power constraint, the transmit power of each antenna is set to a fixed value of $P_o=1$. Each channel coefficient is randomly generated with complex Gaussian distribution of $\Cc\Nc(0,1)$.

\begin{figure}[!t]
  \centering
  \includegraphics[width=3.5in]{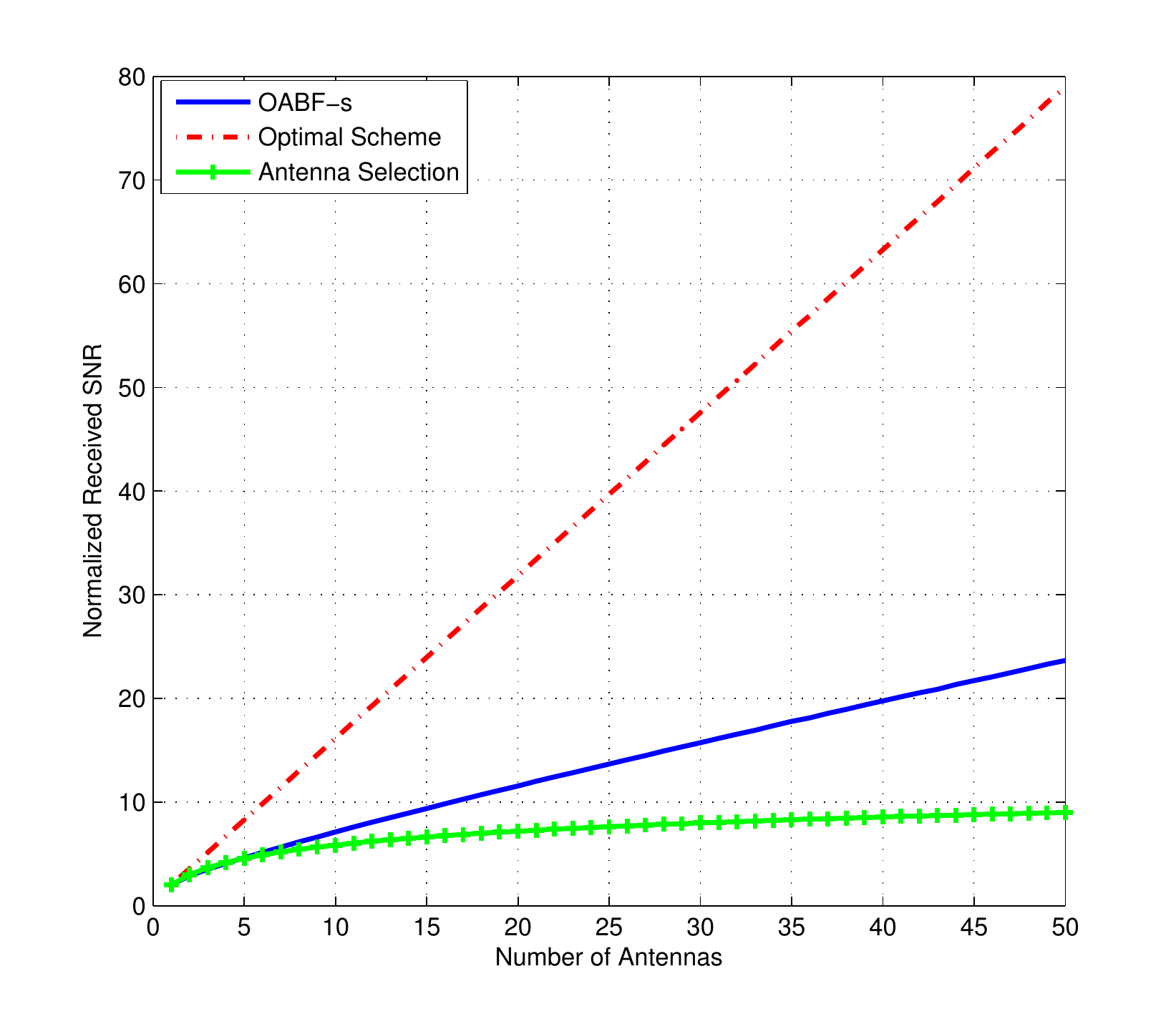}
  \caption{The normalized received SNR in terms of the number of antennas.}
  \label{fig_receive_snr}
\end{figure}

In order to show the array gain of different schemes, we simulate the normalized SNRs at the receiver by increasing the antenna number. The normalized SNR is defined as the received SNR, $\frac{P_o | \sum_{h_j \in T} h_{j} |^2} { \sigma^2}$, over the transmit power, $P_o | T |$. The results are presented in Fig. \ref{fig_receive_snr}. We can see that the normalized SNR of both OABF-s and the optimal scheme increase linearly with the number of antennas, i.e. achieving the full array gain. On the other hand, the normalized SNR of the antenna selection scheme increases in a $\log$ way.

\begin{figure}[!t]
  \centering
  \includegraphics[width=3.5in]{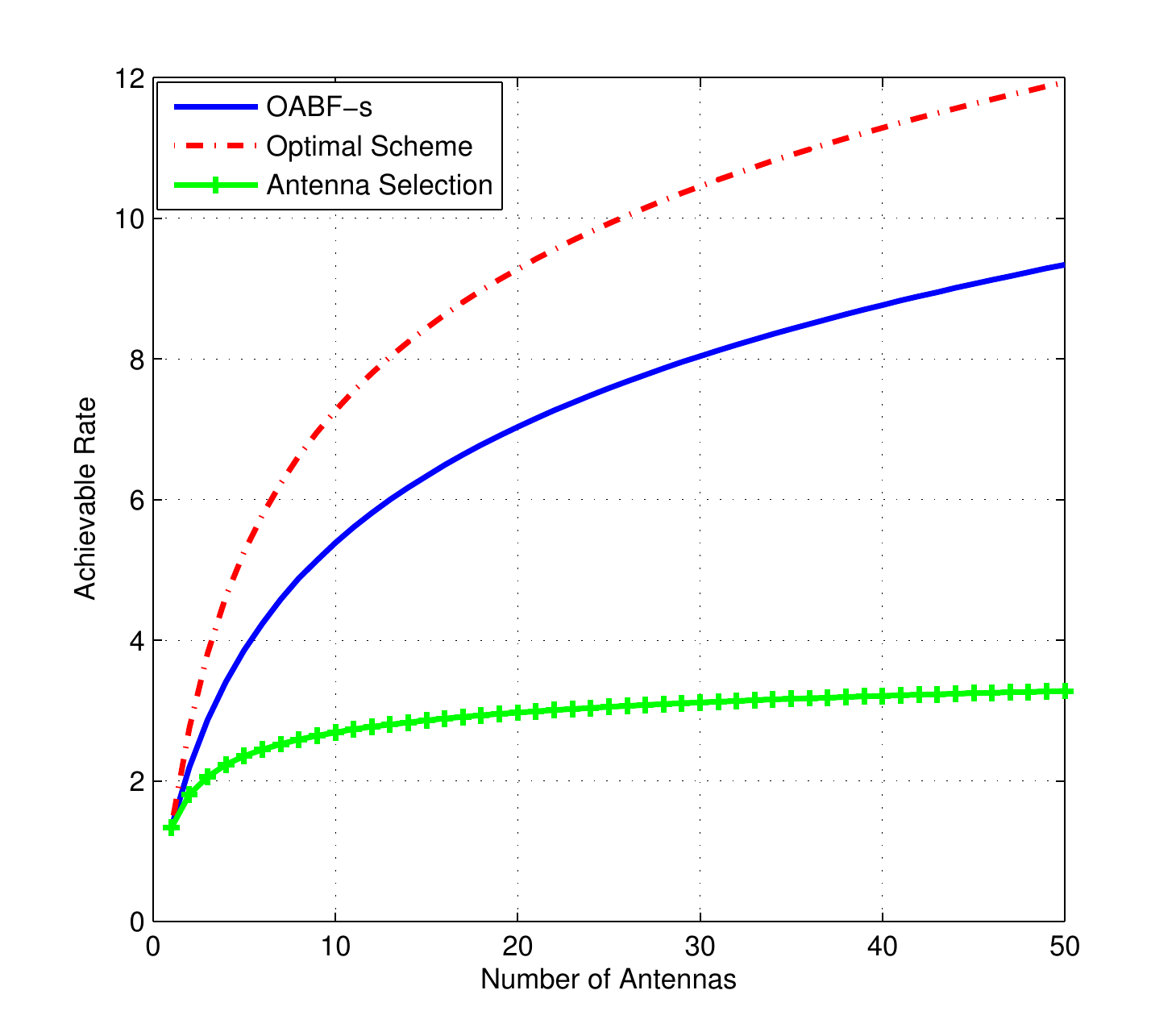}
  \caption{The achievable rate in terms of the number of antennas.}
  \label{fig_achievable_rate_sep}
\end{figure}

In Fig. \ref{fig_achievable_rate_sep},  the average achievable rate at the receiver is obtained by averaging over the instantaneous rates for each channel realizations according to $\log(1+SNR_s)$. We can see that the rates of all three schemes increase with the number of antennas. The gap between OABF-s and the optimal scheme is upper bounded by a constant as stated in Corollary 2a.

\begin{figure}[!t]
  \centering
  \includegraphics[width=\linewidth]{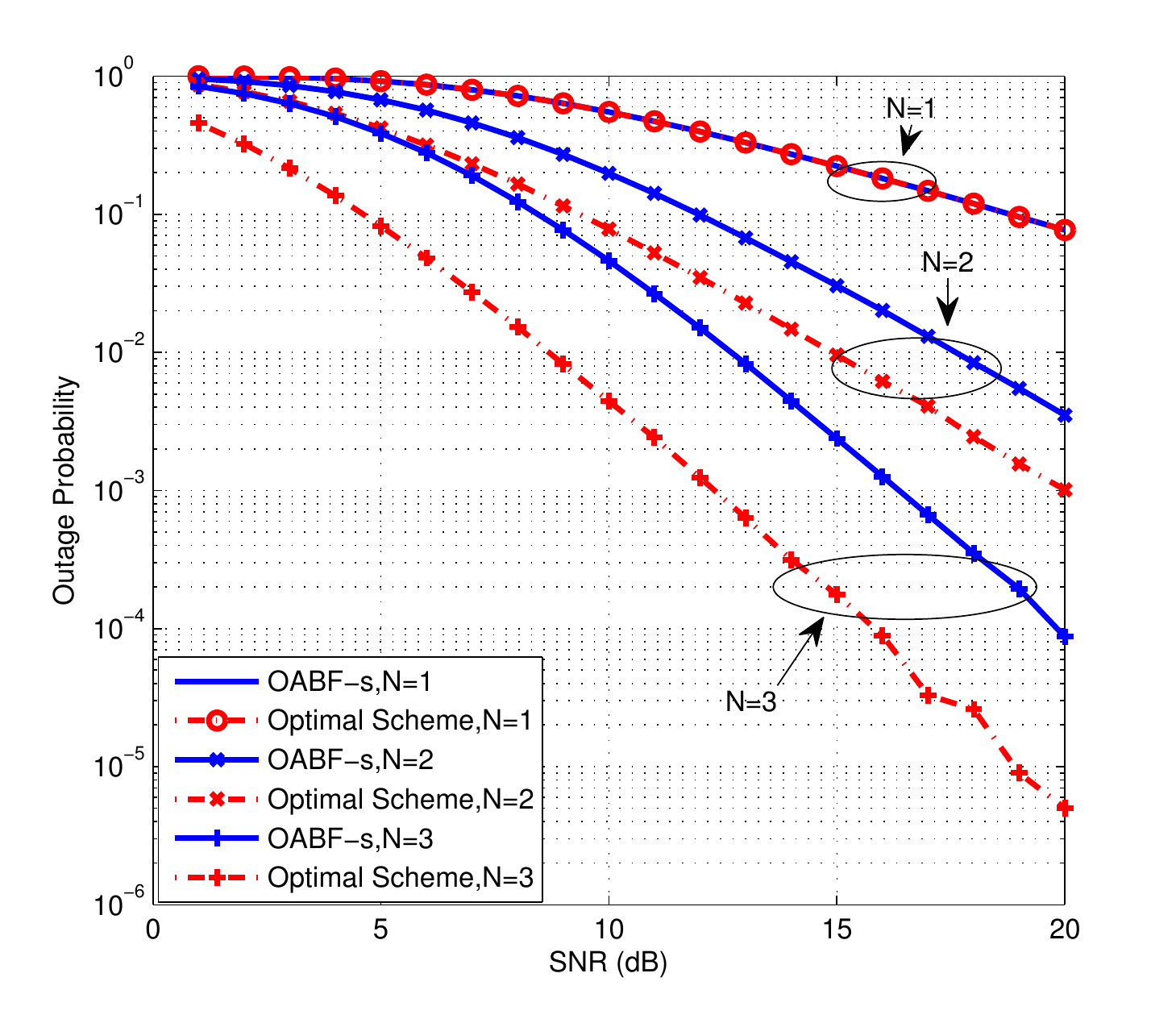}
  \caption{The outage probability in terms of SNR.}
  \label{fig_diversity_sep}
\end{figure}

Fig. \ref{fig_diversity_sep} presents the outage probability, i.e., the probability that the exact received SNR is less then a given threshold. We can clearly see the diversity orders of 1, 2, 3 when the antenna number is $N=1,2,3$, respectively. When $N=1$,  OABF-s and the optimal scheme have the same performance; when $N=2$, there is a gap about 2.5 dB, and this gap increases to 4 dB when $N=3$. 

\begin{figure}[!t]
	\centering
	\includegraphics[width=3.5in]{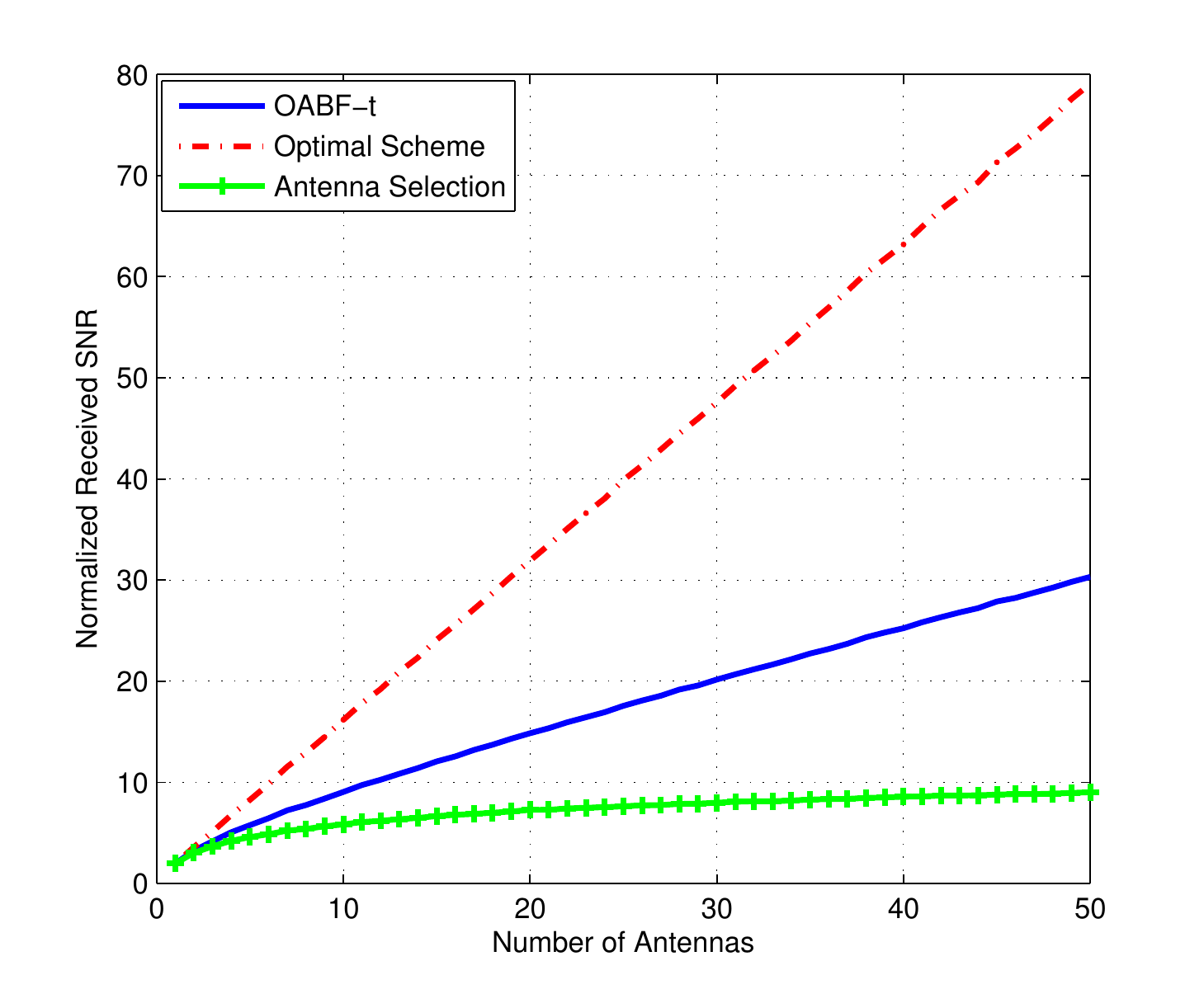}
	\caption{THE received SNR in terms of number of antennas.}
	\label{fig_receive_snr_tot}
\end{figure}

\begin{figure}[!t]
	\centering
	\includegraphics[width=3.5in]{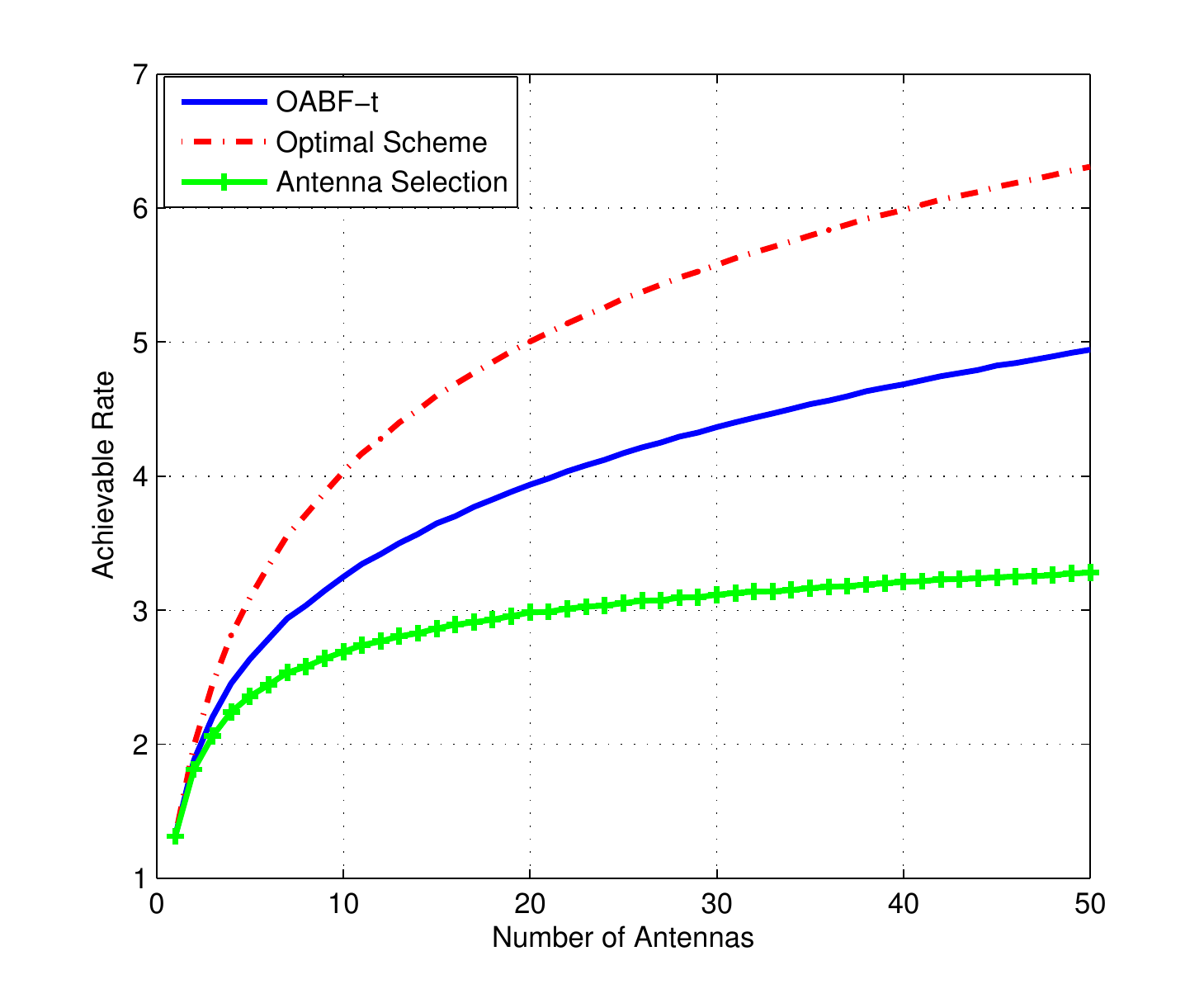}
	\caption{The achievable rate in terms of the number of antennas.}
	\label{fig_achievable_rate_tot}
\end{figure}

\begin{figure}[!t]
	\centering
	\includegraphics[width=3.5in]{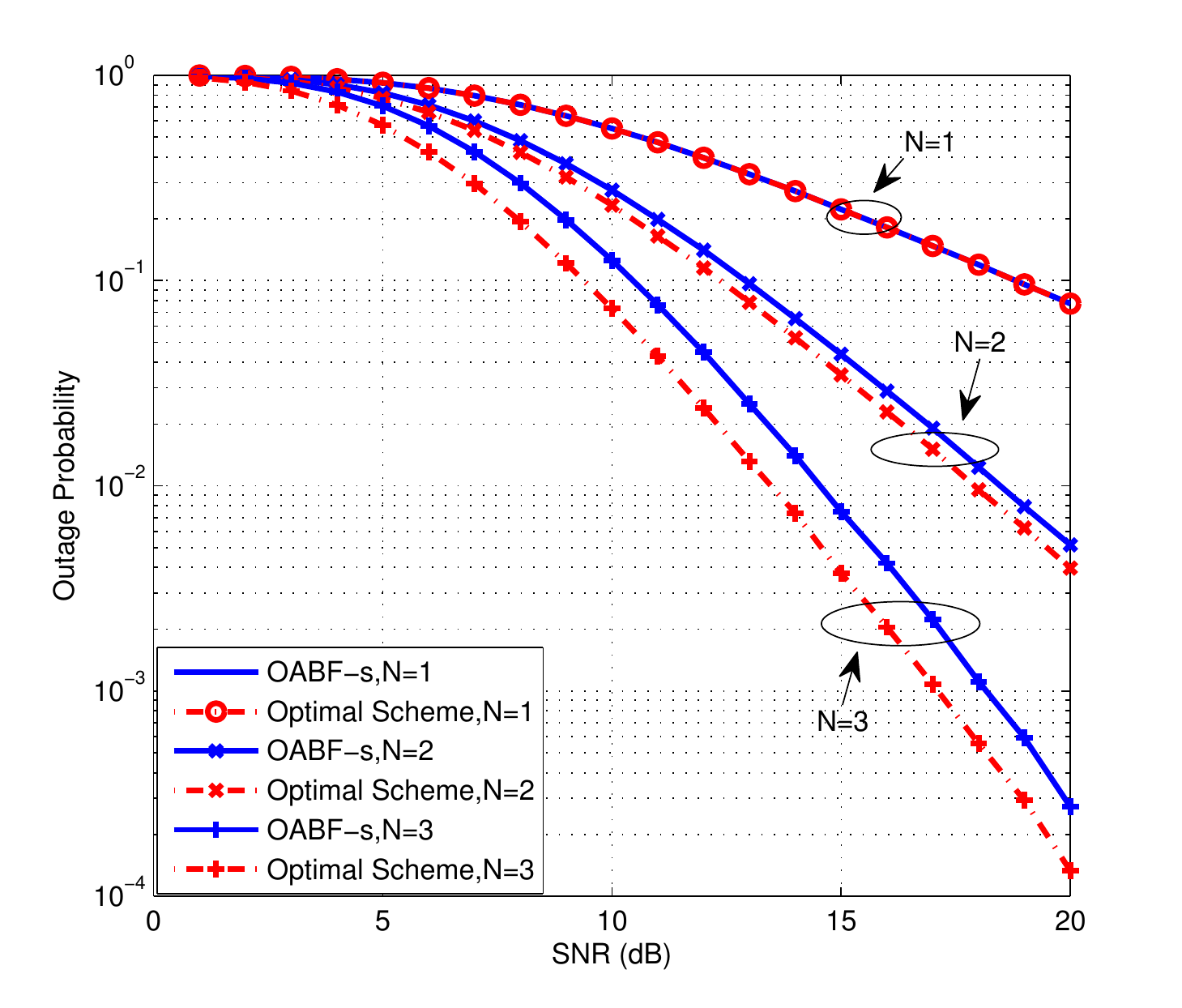}
	\caption{The outage probability in terms of the number of antennas.}
	\label{fig_diversity_tot}
\end{figure}

\subsection{OABF with Total Power Constraint}
We then present the numerical result under the total transmit power constraint $P_t=1$.  In Fig. \ref{fig_receive_snr_tot}, the received SNR with respect to the number of antennas is given. As predicted by Theorem 3, OABF-t can achieve better SNR than OABF-s dose under the per-antenna power constraint. In Fig. \ref{fig_achievable_rate_tot} presents the corresponding achievable rates. We can see that the rate gap between our OABF-t scheme and the optimal scheme is less than 1.5 bits, which is about half of the gap between the OABF-s scheme and the optimal scheme under the per-antenna power constraint.

In Fig. \ref{fig_diversity_tot}, we show the outage probabilities of the optimal scheme and our OABF-t scheme. As we can see that both schemes can achieve the full diversity order of $N$. The gap between the two schemes is about 1 dB for $N=3$, which is much smaller than the case of per-antenna power constraint.

\section{Conclusions and Discussions} \label{sec:conclu}
In this paper, we propose a new analog beamforming architecture, on-off analog beamforming (OABF), which only uses simple analog switches to achieve beamforming gains. To determine the status of each switch with the given channel information, we propose optimal algorithms to maximize the received SNR under the  per-antenna power constraint and total power constraint, respectively.
With linear/polynomial complexities, our OABF can achieve both the full diversity gain and the full array gain. More specifically, the achievable rate gap between the optimal scheme (equal-gain phase-aligned beamforming) and our scheme is a constant of 3.3 bits/symbol, regardless the number of antennas and SNRs.

This paper only discussed the case of one data stream for a point-to-point transmission. Extending it to the case of multiple users and multiple streams is valuable to investigate, both in theoretical analyses and practical algorithm designs. More importantly, our on-off analog beamforming is compatible with other analog beamforming schemes to make up new hybrid architectures. Other beamforming systems, such as Radar, can also adopt our OABF scheme to decrease system cost.

\section *{Appendix: Proof of \it{Theorm 2}}

\begin{IEEEproof} First, it is obvious that the OABF-b algorithm performs better than the antenna selection scheme that can achieve the full diversity gain of $N$. Therefore, both the suboptimal OABF-b and optimal OABF-s algorithms can achieve the full diversity gain.

We now consider the array gain. With per-antenna power constraint, it is easy to see that the phase-aligned analog beamforming scheme is optimal \cite{cophase_combining_miso_2007} to make full use of each antenna's power. For this optimal phase-aligned analog beamforming, the expected total received signal power, averaging over all possible phases, can be calculated as
\beq \label{eq:optimal_scheme}
P_{op}=  \left( \sum_{j=1}^{N} |h_{j}| \right)^2 = \sum_{j=1}^{N} |h_{j}|^2 + 2\sum_{j=1}^{N} \sum_{k=1,k\neq j}^{K_2}  |h_k h_j|.
\eeq

We then show that the expected received signal power of the bridge algorithm, OFBA-b, is larger than $P_{op}/\pi^2$. According to the OFBA- algorithm, 
the expected received signal power is

\begin{align} \label{eq:pbri_definition}
P_{bri} =  \E |\sum_{h_j \in R} h_{j} |^2
=\E |\sum_{j=1,-\pi/2 \leq \theta_j \leq \pi/2}^{N} h_{j} |^2
\end{align}
where $\theta_j$ is the phase difference between $h_j$ and $h_m$. We can decompose the summation vector $\sum_{j=1}^{|R|} h_{j}$
into two components, one align the direction of $h_m$ and the other orthogonal to it. Then, the total power must be larger than or equal to that of the component align the direction of $h_m$. Therefore, we have
\begin{align} \label{eq:pbri_component}
P_{bri} &\geq \E \left( \sum_{j=1,-\pi/2 \leq \theta_j \leq \pi/2}^{N} |h_{j}|\cos(\theta_j) \right)^2 .
\end{align}

On the other hand, all channel coefficients, $h_j$, $j=1,2,\cdots, N$, are assumed to be independent Gaussian random variables, and thus their phases are uniformly distributed between $[0,2\pi]$. Since our bridge algorithm always selects $h_j$ on the same half plane as $h_m$, then we have $\E |\sum_{h_j \in R} h_{j} |^2 \geq  \E |\sum_{h_j \notin R} h_{j} |^2 $. With respect to \eqref{eq:pbri_definition} and \eqref{eq:pbri_component}, we can obtain the following inequalities as
\begin{align}
&P_{bri} \geq \frac{1}{2}\E |\sum_{h_j \in R} h_{j} |^2 +\frac{1}{2}\E |\sum_{h_j \notin R} h_{j} |^2 \nonumber \\
&\geq \frac{1}{2}\E \left( \sum_{-\pi/2 \leq \theta_j \leq \pi/2} |h_{j}|\cos(\theta_j) \right)^2 +  \nonumber \\
          & \qquad \qquad \qquad \frac{1}{2}\E \left( \sum_{\pi/2 \leq \theta_j \leq 3\pi/2} |h_{j}|\cos(\theta_j) \right)^2 \nonumber \\
&\geq \frac{1}{4}\E \left( \sum_{-\pi/2 \leq \theta_j \leq \pi/2} |h_{j}|\cos(\theta_j) -  \sum_{\pi/2 \leq \theta_j \leq 3\pi/2} |h_{j}|\cos(\theta_j) \right)^2 \nonumber \\
&= \frac{1}{4}\E \left( \sum_{j=1}^{N} |h_{j}\cos(\theta_j)| \right)^2
\end{align}
where the last inequality comes from the fact that $2a^2+2b^2 \geq (a+b)^2$.  We can then calculate the expectation by averaging over all phases as
\begin{align} \label{eq:performance_difference_oabfs}
&P_{bri}
\geq \frac{1}{4} \sum_{j=1}^{N} \frac{2}{\pi} \int_{0}^{\pi/2} |h_{j}|^2\cos^2(\theta_j) \,d\theta_j + \nonumber \\
&\qquad  \frac{2}{4}\sum_{j=1}^{N} \sum_{k=1,k\neq j}^{N}  \frac{4}{\pi^2}\int_0^{\pi/2} \int_0^{\pi/2} |h_k h_j| \cos(\theta_j)\cos(\theta_k)\,d\theta_j\,d\theta_k \nonumber \\
&=\frac{1}{4} \left(\frac{1}{2}  \sum_{j=1}^{N}  |h_{j}|^2 + 2  \frac{4}{\pi^2} \sum_{j=1}^{N} \sum_{k=1,k\neq j}^{N} |h_k h_j| \right)\nonumber \\
&\geq \frac{1}{4}\frac{4}{\pi^2}  \left( \sum_{j=1}^{N}  |h_{j}|^2 + 2  \sum_{j=1}^{N} \sum_{k=1,k\neq j}^{N} |h_k h_j| \right)\nonumber \\
&= \frac{1}{\pi^2} P_{op}
\end{align}

By further averaging over all the amplitude in \eqref{eq:optimal_scheme}, the expected SNR of the optimal phase-aligned beamforming scheme can be obtained as $\left( N+N(N-1)\pi /4 \right)$ \cite{linear_diversity_1959}. Then, the expected $SNR_s$ is lower bounded by
\begin{align} \label{eq:gamma_1_result}
\gamma_1 &= \E \left[ SNR_s \right]\geq \frac{P_o}{\pi^2 \sigma^2} \E P_{op} \nonumber \\
&= \frac{P_o}{\pi^2 \sigma^2} \left( N+N(N-1)\pi /4 \right)
\end{align}
The ratio of $\gamma_1$ over the SNR of a single-input single-output system, which is  $\frac{NP_o}{\sigma^2}$, is of the order of $N$. Therefore, our OABF-b algorithm can also achieve the full array gain of $N$.
\end{IEEEproof}




\bibliographystyle{IEEEtran}
\bibliography{final_refs}


\end{document}